\documentclass[aps,prl,groupedaddress,reprint]{revtex4-1}
\usepackage{amsmath}
\usepackage{amsfonts}
\usepackage{graphicx}
\usepackage{lipsum}

\begin{document}

\title{Pressure and temperature dependence of solubility and surface adsorption of nitrogen in the liquid hydrocarbon bodies on Titan}

\author{Pradeep Kumar}
\email[To whom correspondence should be addressed. Email: ]{pradeepk@uark.edu}
\affiliation{Department of Physics, University of Arkansas, Fayetteville, AR, 72701}
\affiliation{Arkansas Center for Space and Planetary Sciences, University of Arkansas, Fayetteville, AR, 72701.}
\author{Vincent F. Chevrier}
\affiliation{Arkansas Center for Space and Planetary Sciences, University of Arkansas, Fayetteville, AR, 72701.}

\date{\today}

\begin{abstract}
We have studied the pressure and temperature dependence of solubility of nitrogen in methane and ethane using vapor-liquid equilibrium simulations of binary mixtures of nitrogen in methane and ethane for a range of pressures between $1.5$~atm and $3.5$~atm and temperatures between $90$~K and $110$~K, thermodynamic conditions that may exist on the Saturn's moon, Titan. We find that the solubility of nitrogen in methane increases linearly with pressure while the solubility of nitrogen in ethane increases exponentially with pressure at temperature $90$~K. Solubility of nitrogen in both methane and ethane exhibits an exponential decrease with temperature at a pressure of $3$~atm. The solubility of nitrogen in methane is much larger compared to that in ethane in the range of pressure and temperature studied here.  Our results are in quantitative agreement with the available experimental measurements of the solubility of nitrogen in methane and ethane. Furthermore, we find that the surface adsorption of nitrogen increases with increasing pressure at temperature $90$~K, while the adsorption free energy increases with increasing pressure.  Moreover, we find that the surface tension decreases linearly with pressure for both nitrogen-methane and nitrogen-ethane systems. The rate of decrease of surface tension with pressure for nitrogen-ethane system is much larger as compared to the nitrogen-methane system. Finally, we find that the absorption of a nitrogen molecule into the liquid-phase from the interface is diffusive and does not involve any appreciable energy barrier. Our results suggest that homogeneous nucleation of bubbles is unlikely on Titan and the bubble formation in the lakes on Titan must arise from heterogeneous nucleation of bubbles.
\end{abstract}


\maketitle

Saturn's giant moon, Titan, holds a unique place in our solar system. It is the only other planetary body, besides Earth, that possesses  stable liquid on its surface~\cite{Lunine:2008aa,Lorenz:2010aa,Hayes:2018aa,Lorenz:2010aa,Hayes:2018aa}. Similar to Earth, the atmosphere of Titan is rich in nitrogen. The primary constituents of Titan's atmosphere are nitrogen, methane, and ethane. Due to low surface temperature and pressure about $1.5$~atm~\cite{Lunine:2008aa,Lorenz:2010aa,Hayes:2018aa}, both methane and ethane condense out of the atmosphere and exist in forms of lakes and seas on the surface in dynamic equilibrium with its atmosphere~\cite{Lorenz:2010aa} giving rise to a hydrological cycle similar to earth. Titan exhibits the richest chemistries in our solar system. Major neutral species such as nitrogen and methane undergo photo-dissociation giving rise to simple and complex organic molecules. The Infrared Radiometer Interferometer and Spectrometer onboard Voyager revealed the presence of methane (CH$_4$), molecular hydrogen (H$_2$), ethane (C$_2$H$_6$), acetylene (C$_2$H$_2$), ethylene (C$_2$H$_4$), hydrogen cyanide (HCN), and carbon dioxide (CO$_2$)~\cite{Flasar:1983aa,Toon:1988aa,Lunine:1983aa,Brown:2008aa}. Earth based telescopes later discovered new molecules including acetonitrile (CH$_3$CN), carbon monoxide (CO), and water (H$_2$O). Cassini mission to Saturn has further revealed the presence of other molecules such as ammonia (NH$_3$), vinyl cyanide (C$_2$H$_3$CN).

Recent experiments, thermodynamics models, and molecular simulations suggest that the liquid hydrocarbon bodies on the surface are major sinks of nitrogen~\cite{MALASKA201794,farnsworth2019nitrogen,Cheverier:2018aa,kumar2020titan}. Most of these experiments were motivated by the detection of transient bright features by Cassini RADAR system, also known as "Magic Islands"~\cite{Hofgartner:2014aa,HOFGARTNER2016338}. It was hypothesized that these bright features arise due to the exsolution of nitrogen from the lakes in the form of bubbles that could be detected by Cassini~\cite{Cordier:2017aa,cordier2018bubbles}. The formation of bubble in the lakes could be driven by superheating or supersaturation of binary mixtures of hydrocarbon liquid and nitrogen gas. Indeed, recent studies on solubility of nitrogen in methane, ethane, and a mixture of methane and ethane in different proportions suggest that nitrogen is highly soluble in methane and the solubility exhibits a sharp temperature dependence~\cite{Llave:1985aa,Llave:1987aa,Cordier:2017aa,MALASKA201794,Cheverier:2018aa}. Various scenarios for the exsolution of nitrogen in the form of bubbles were hypothesized, including superheating, supersaturation, titration of ethane into methane rich regions~\cite{MALASKA201794,cordier2018bubbles}. Explicit experiments on the bubble formation in Titan-like conditions were recently performed and authors found several criteria for the formation of bubbles on Titan~\cite{farnsworth2019nitrogen}. The focus of many of these experiments have solely been the 
 solubility of nitrogen and exsolution of nitrogen from the lakes and seas of Titan~\cite{MALASKA201794,farnsworth2019nitrogen}. Theoretical work using thermodynamic models utilizing available experimental data have focused on the same process~\cite{cordier2009,Cordier:2017aa,cordier2018bubbles}. We have recently investigated the phase-equilibria of nitrogen and hydrocarbon liquids on Titan using vapor-liquid equilibrium simulations at a pressure of $1.5$~atm and temperatures between $90$~K and $110$~K~\cite{kumar2020titan}. We find that the solubility of nitrogen decreases with increasing temperature and the the solubilities values are in quantitative agreement with available experimental data. In addition to increase of solubility upon decreasing temperature, we find that the surface tension of the gas-liquid interface decreases with temperature. Moreover, the surface tension of the nitrogen-methane binary mixture is about half the value of the surface tension of the nitrogen-ethane binary mixture at the same temperature at a pressure of $1.5$~atm, suggesting that it would be easier to form a critical nucleus of a bubble in the nitrogen-methane mixture as compared to nitrogen-ethane mixture. Furthermore, in these studied we found a strong temperature-dependent surface adsorption where nitrogen forms a dense surface layer of about $2$~nm thickness~\cite{kumar2020titan}. The complex and rich chemistries on Titan with the presence of liquid hydrocarbon lakes on the surface poses many challenging questions. Many of compounds in the atmosphere condense out of atmosphere and precipitate onto the liquid hydrocarbon bodies on the surface of Titan~\cite{kasting2006atmospheric,lorenz2008titan,cable2012titan,horst2017titan}. As they precipitate on the surface they will first interact with this dense layer of nitrogen instead of the bulk  liquid hydrocarbon. Therefore, it is also important to understand how the surface adsorption of nitrogen changes with pressure and temperature. Cassini CIRS (Composite Infrared Spectrometer) observation of Titan's surface temperature between $2004$ and $2016$ suggest that the surface temperature of the Titan varied between $~90$K and $95$~K~\cite{jennings2016surface}. Cassini data has further revealed the depth of liquid hydrocarbon lakes and seas on Titan. Kraken Mare, the largest hydrocarbon sea on Titan is at least $35$~m deep~\cite{mastrogiuseppe2019deep}. The maximum depth of Ligeia Mare have been estimated to be between $100$~m and $200$~m~\cite{mastrogiuseppe2014bathymetry,le2016composition}, making the static pressure at the bottom of this lake to be about $\approx 2.7$~atm, assuming a pure methane body.

In order to investigate the pressure and temperature dependence of the solubility and surface adsorption of nitrogen in methane and ethane, we have performed molecular dynamics vapor-liquid-equilibrium (VLE) simulations~\cite{buldyrev2007water,Minkara:2018aa,panagiotopoulos2002direct,panagiotopoulos1987direct,potoff2001vapor,morrow2019vapor} of binary mixtures of nitrogen and methane (nitrogen-methane) and nitrogen and ethane (nitrogen-ethane) at pressures between $1.5$~atm and $3.5$~atm and temperatures between $90$~K and $110$~K, thermodynamic conditions that may exist on Titan. 

In order to investigate the pressure and temperature dependence of the solubility and surface adsorption of nitrogen in methane and ethane, we have performed molecular dynamics vapor-liquid-equilibrium (VLE) simulations~\cite{buldyrev2007water,Minkara:2018aa,panagiotopoulos2002direct,panagiotopoulos1987direct,potoff2001vapor,morrow2019vapor} of binary mixtures of nitrogen and methane (nitrogen-methane) and nitrogen and ethane (nitrogen-ethane) at pressures between $1.5$~atm and $3.5$~atm and temperatures between $90$~K and $110$~K, thermodynamic conditions that may exist on Titan. 

\section*{Results}
\subsection* {Pressure and temperature dependence of the solubility of nitrogen in methane and ethane}
\medskip
\begin{figure}
\begin{center}
\includegraphics[width=0.8\linewidth]{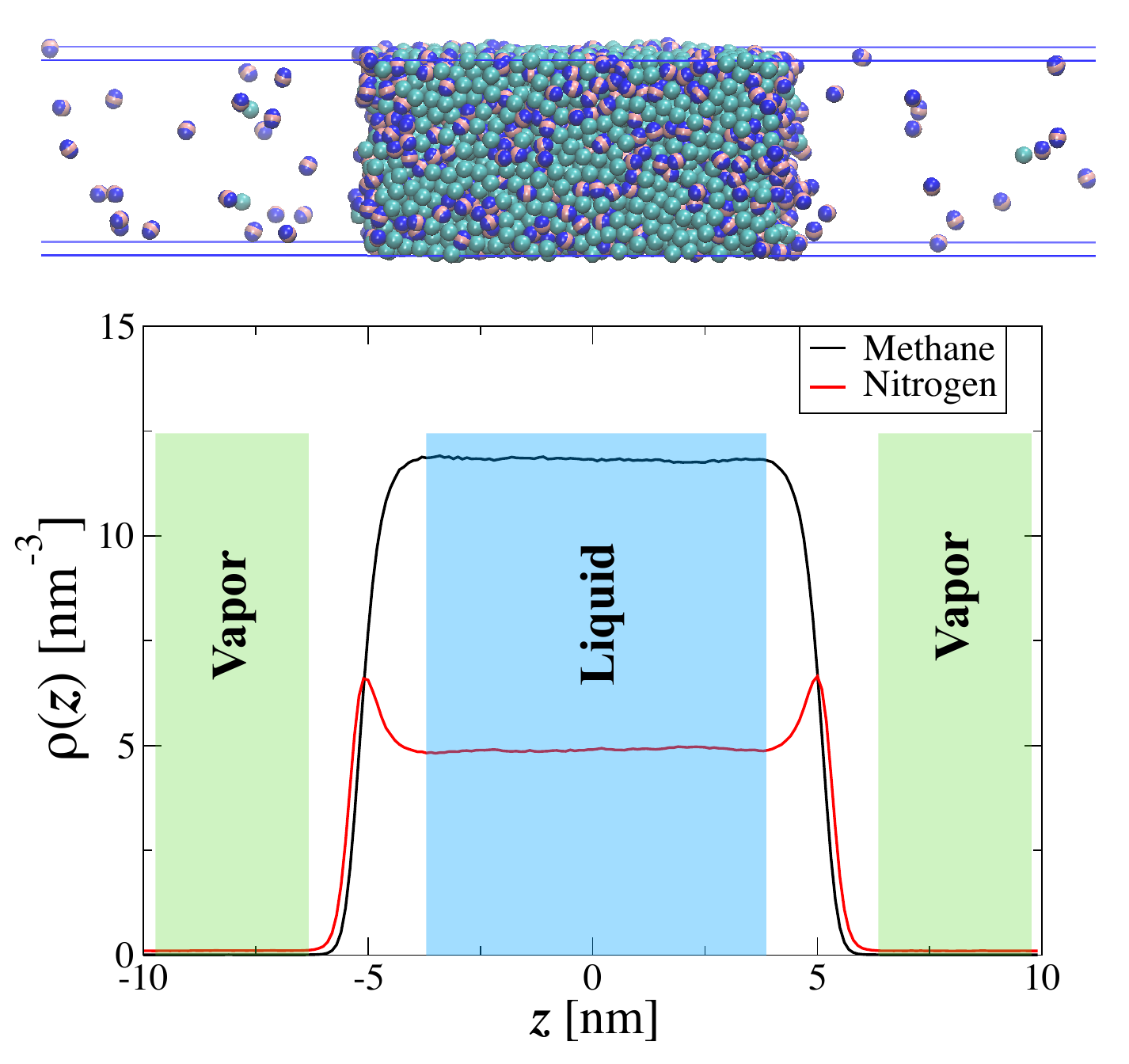}
\end{center}
\caption{(Top) A snapshot of the nitrogen-ethane binary mixture at $T=90$K. (Bottom) Average number density profile, $\rho(z)$,  for ethane and nitrogen along the $z$-direction. The blue shaded region represents the liquid phase and the green shaded region represents the vapor phase.}
\label{fig:fig1}
\end{figure}
\begin{figure}
\centering
\includegraphics[width=0.98\linewidth]{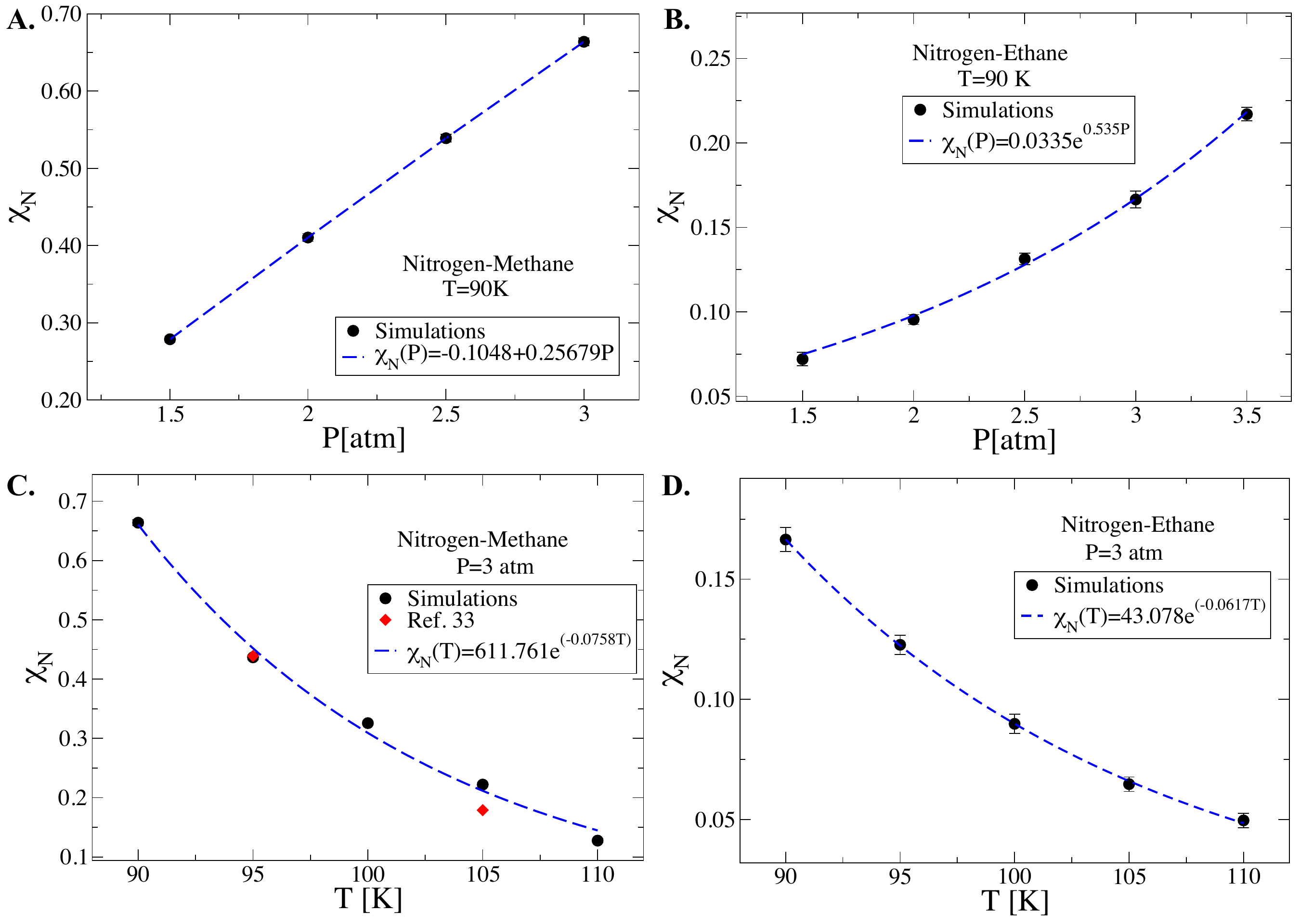}
\caption{(A) Mole-fraction, $\chi_N$, of the nitrogen in methane as a function of pressure at $T=90$~K. The dotted line is a linear fit, $-0.1048+0.25679P$, through the data. The solubility of nitrogen in methane increases linearly with pressure.  (B) Mole-fraction, $\chi_N$, of nitrogen in ethane as a function of pressure at $T=90$~K. In contrast to the solubility of nitrogen in methane, the solubility of nitrogen in ethane exhibits an exponential dependence on pressure and can be fit well with  $\chi_N(P)=0.0335e^{0.535P}$. (C) Mole-fraction, $\chi_N$, of nitrogen as a function of temperature for $P=3.0$~atm. The solubility of nitrogen in methane decreases exponentially with temperature and and be fit with  $\chi_N(T)=611.761e^{{-0.0758T}}$. For a comparison, we also show two experimental data points (in red diamonds) from Ref.~\cite{baidakov2011capillary}. The simulation data agree well with the experimental values. (D) Mole-fraction, $\chi_N$, of nitrogen in ethane as a function of temperature for $P=3.0$~atm. The solubility of nitrogen in ethane also decreases exponentially with temperature and can be fit well with  $\chi_N(T)=43.07e^{{-0.0617T}}$. The error in the solubility is estimated from the errors in the mole fractions of hydrocarbon and nitrogen in the liquid phase.}
\label{fig:fig2}
\end{figure}

Typical average number density profile, $\rho(z)$, for nitrogen-methane binary mixture in equilibrium at temperature $T=90$~K and at pressure $P=1.5$~atm is shown in Figure~\ref{fig:fig1}. The low density vapor phase coexist with the high density liquid phase. To avoid the interface, we define the liquid phase (or the gas phase) as the region where the $z$-derivative of the density of methane/ethane and nitrogen is zero (ses Fig.~\ref{fig:fig2}). The solubility is defined as the average mole-fraction of nitrogen in the liquid phase of binary mixtures at equilibrium.  We count the number of molecules of nitrogen, $\mathcal{N}_N^{\ell}$, in the liquid phase of methane/ethane and similarly count the number of molecules of methane/ethane, $\mathcal{N}_M^{\ell}$ or $\mathcal{N}_E^{\ell}$, in the liquid-phase for the corresponding binary mixture simulations. The solubility as measured by the mole-fraction, $\chi_N$, of nitrogen in the liquid phase, and is defined as
\begin{align}
\chi_{N} &= \frac{\mathcal{N}_N^{\ell}}{\left(\mathcal{N}_N^{\ell}+N_M^{\ell}\right)} \text{ For nitrogen-methane binary system} \\
	     &= \frac{\mathcal{N}_N^{\ell}}{\left(\mathcal{N}_N^{\ell}+N_E^{\ell}\right)}  \text{ For nitrogen-ethane binary system}
\end{align}

In Figs.~\ref{fig:fig2}(A) and (B), we show the mole-fraction of methane, $\chi_N$, as a function of pressure for nitrogen-methane and nitrogen-ethane binary systems at $T=90$~K, respectively. The error in solubility is estimated from the error on the mole-fractions of hydrocarbon and nitrogen in the liquid-phase. Solubility of nitrogen increases with pressure for both the systems. Moreover, the solubility of nitrogen in methane increases linearly with pressure while the solubility in ethane exhibits an exponential dependence on pressure and be fit well with $\chi_N(P)=0.0335e^{0.535P}$. The temperature dependence of the solubility at a fixed pressure of $P=3.0$~atm for nitrogen-methane and nitrogen-ethane binary systems are shown in Figs.~\ref{fig:fig2}(C) and (D), respectively.  Similar to earlier work for $P=1.5$~atm~\cite{kumar2020titan}, the solubility decreases exponentially with increasing temperature for $P=3.0$~atm and can be fit with  $\chi_N(T)=611.761e^{{-0.0758T}}$ for nitrogen in methane,  and $\chi_N(T)=43.078e^{{-0.0617T}}$ for nitrogen in ethane. For a comparison, we also show the available experimental data of solubility of nitrogen in methane from Ref.~\cite{baidakov2011capillary}. The simulation results of the solubility agree well with the experimental values.
\subsection*{Pressure and temperature dependence of surface tension}

\begin{figure}
\centering
\includegraphics[width=0.98\linewidth]{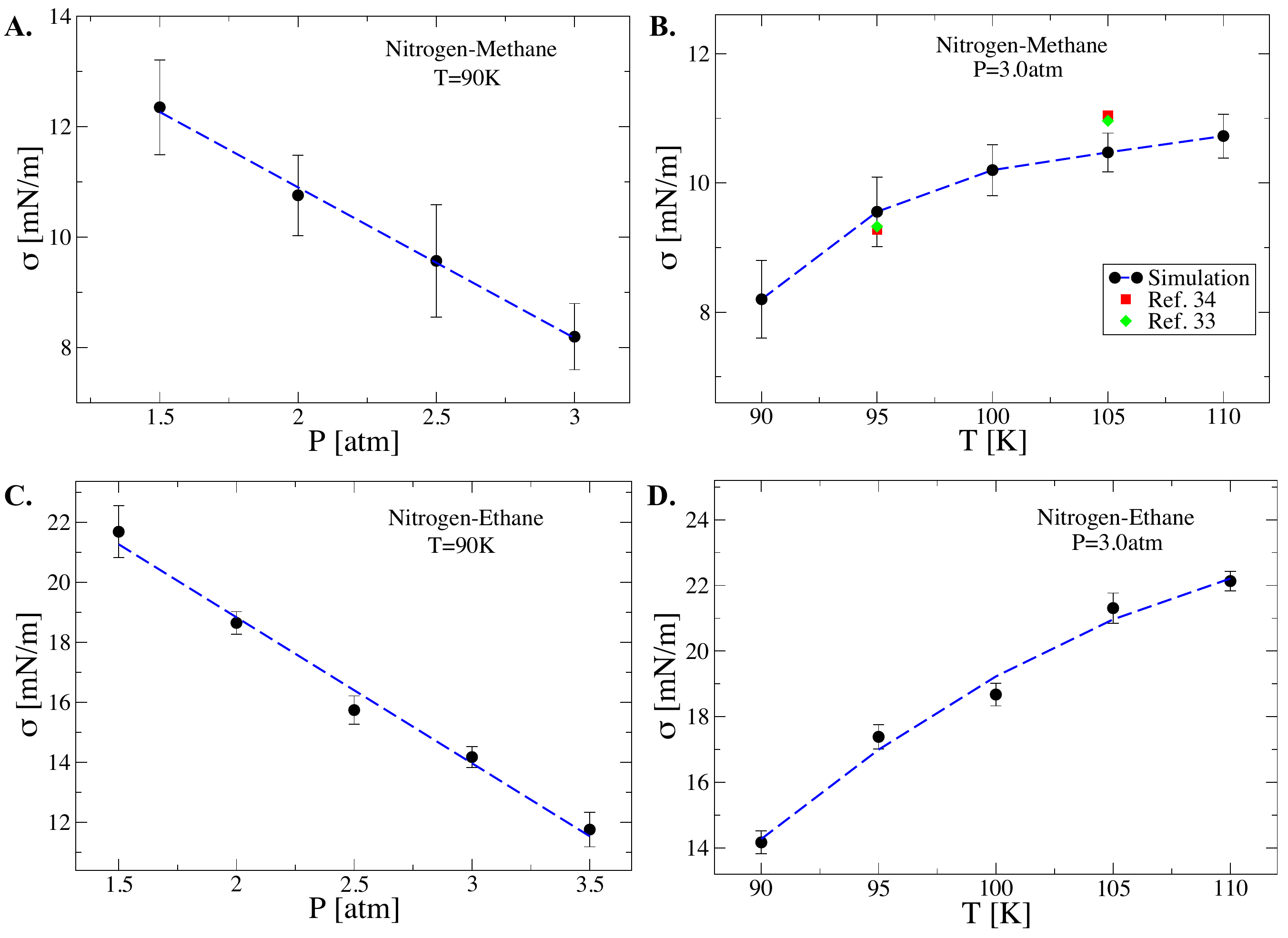}
\caption{Pressure dependence of the surface tension, $\sigma$, for (A) nitrogen-methane and (C) nitrogen-ethane systems at $T=90$~K. Surface tension for the nitrogen-ethane system is about twice as large compared to surface tension of the nitrogen-ethane system in the range of temperature and pressure studied. Moreover, the surface tension for both the nitrogen-methane and nitrogen-ethane system decreases linearly with pressure. Dotted lines are the linear fits $\sigma = 16.355-2.727P$ and $\sigma = 28.572-4.869P$ through the nitrogen-methane and nitrogen-ethane data points, respectively. Temperature dependence of the surface tension for nitrogen-methane and nitrogen-ethane systems at $P=3$~atm are shown in (B) and (D), respectively. The dotted lines are guide to the eye. Surface tension for both the systems increases upon increasing temperature.}
\label{fig:fig3}
\end{figure}

 We next studied the behavior of surface tension for the thermodynamic conditions studied here, which is easily available from the molecular dynamics simulations. The surface tension, $\sigma$, is defined as
\begin{equation}
\sigma = \frac{L_z}{2}\left[P_{zz} -0.5(P_{xx}+P_{yy})\right]
\end{equation}
where $L_z$ is the box-length in the $z$-direction and $P_{xx}$, $P_{yy}$, $P_{zz}$ are the diagonal components of the pressure tensor in the $x$, $y$, and $z$-directions, respectively. A factor of $2$ accounts for the presence of two interfaces in the simulation box. Figures ~\ref{fig:fig3} (A) and (C) show the pressure dependence of the surface tension, $\sigma$, at $T=90$~K for the nitrogen-methane and the nitrogen-ethane binary mixtures, respectively.  Surface tension decreases linearly with pressure for both the systems.  Furthermore, we find that the rate of surface tension decrease with pressure, $|\left(\frac{d\sigma}{dP}\right)_T|$, is larger for nitrogen-ethane system compared to nitrogen-methane. The dotted lines in Figs.~\ref{fig:fig3}(A) and (C) are the linear fits $\sigma = 16.355-2.727P$ and $\sigma = 28.572-4.869P$ through the nitrogen-methane and nitrogen-ethane data points, respectively. In Figs.~\ref{fig:fig3}(B) and (D), we show the temperature dependence of surface tension at $P=3.0$~atm for nitrogen-methane and nitrogen-ethane systems, respectively. Surface tension increases with increasing temperature for both the systems. For a comparison, we also show experimental data (solid red squares and green diamonds) from Refs.~\cite{LEMMON-RP10,baidakov2011capillary} Our results for surface tension values are in quantitative agreement with the experimental data~\cite{baidakov2011capillary,baidakov2016surface,blagoi1960surface}. Unlike the linear behavior of $\sigma$ with pressure, we find a non-linear dependence of $\sigma$ on temperature.
Surface tension plays an important role in bubble nucleation and growth. According to classical nucleation theory, the average rate of formation of critical nuclei, $J$, per unit solution volume per unit time is given by~\cite{volmer1939kinetik,becker1935kinetische,zeldovich1946acta,frenkel1939general} 
\begin{equation}
J = \sqrt{\frac{2\sigma}{\pi m_{N_2}}}\rho_\ell \rm{exp}\left(-\frac{4\pi\sigma  {r_c}^2}{3k_BT}\right)
\label{eq:eqJ}
\end{equation}
where $\sigma$ is the surface tension, $m_{N_2}$ is the mass of single nitrogen molecule, $\rho_\ell$ is the number density of nitrogen in the liquid, $k_B$ is the Boltzmann constant, $T$ is the temperature, and $r_c$ is the critical radius of the bubble and is given by
\begin{equation}
r_c = \frac{2\sigma}{p_{g}-p_\ell}
\end{equation}
where $p_{g}$ is the pressure of the gas in the critical bubble, and $p_\ell$ is the pressure of the liquid. We use Eq.~\ref{eq:eqJ} to estimate the nucleation rate upon supersaturation of nitrogen in the liquid phase. Let's consider an extreme scenario in which the liquid is saturated with nitrogen at $P=3.0$~atm and $T=90$~K and suddenly the pressure drops to $1.5$~atm and temperature rises to $T=100$~K, i.e. there is a decompression of $1.5$~atm and temperature rises by $10$~K. Let's further assume that there is no degassing and the system instantly comes to the new pressure and temperature. The liquification-pressure of nitrogen is  $7.92$~atm at $T=100$~K~\cite{span2000reference}. Let's assume the most favorable condition for bubble nucleation by assuming the bubble to be composed entirely of nitrogen and the pressure inside the bubble to be $7.92$~atm, the maximum pressure the bubble can have. Now Eq.~\ref{eq:eqJ} can then be used to estimate the average rate of nucleation of the supersaturated solution. If we use the value of $\sigma=8.2mN/m$, the surface tension corresponding to $P=3$~atm and $T=90$~K, we find $r_c=25.46$~nm, and an extremely small value of $J$ with $log(J)=-1.453\rm{x}10^4$. Such a small value of $J$ suggests that homogeneous nucleation is unlikely in Titan's liquid hydrocarbon bodies even under an extreme scenario considered here. Other considerations such as supersaturated  ethane or a mixture of methane and ethane do not change this either. Similar estimates have been made about the homogeneous nucleation rate by Cordier et. al.~\cite{cordier2018bubbles}. The likely scenario of the bubble formation on the Titan can be heterogeneous nucleation whereas the surfaces decrease the energy required to form a critical bubble~\cite{frenkel1955kinetic}. 
\begin{figure}
\centering
\includegraphics[width=0.98\linewidth]{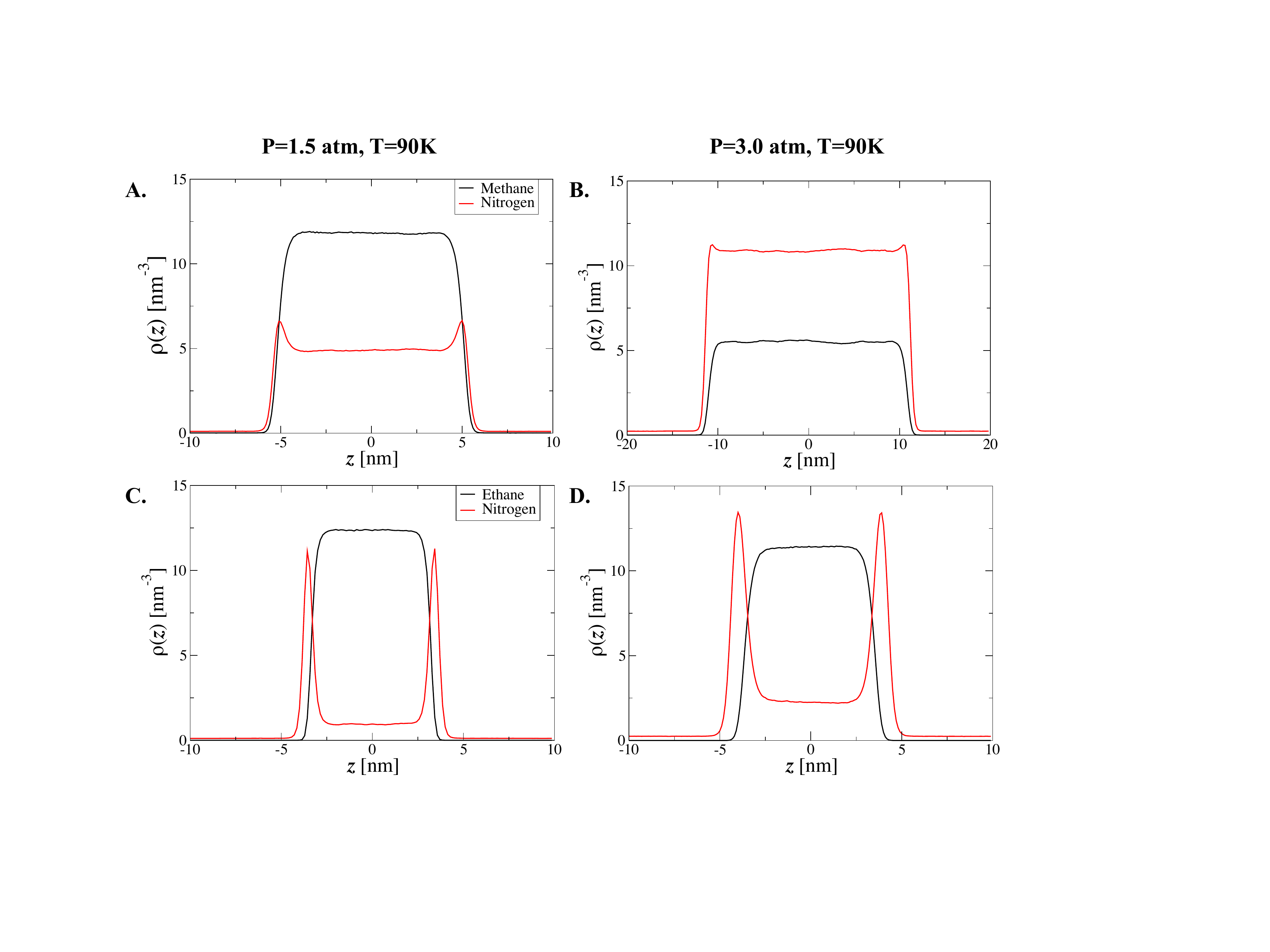}
\caption{Number density profile, $\rho(z)$, at a fixed temperature $T=90$~K for (A) nitrogen-methane system at $P=1.5$~atm, (B) nitrogen-methane system at $P=3.0$~atm. The density of the nitrogen in liquid-phase as well as at the interface increases upon increasing pressure.Number density profile, $\rho(z)$, at a fixed temperature $T=90$~K for (C) nitrogen-ethane system at $P=1.5$~atm, (D) nitrogen-ethane system at $P=3.0$~atm. The density of the nitrogen in liquid-phase as well as at the interface increases upon increasing pressure. }
\label{fig:fig4}
\end{figure}
\subsection*{Adsorption of nitrogen at the interface}
In our recent work, we investigated the solubility and surface adsorption of nitrogen in binary and ternary mixtures of methane/ethane/nitrogen~\cite{kumar2020titan}. We find a strong temperature-dependent adsorption of nitrogen on the surface for $P=1.5$~atm.  In this section, we investigate how the pressure affects the surface adsorption of nitrogen at the nitrogen-methane and the nitrogen-ethane interface. In Figs.~\ref{fig:fig4} (A) and (B), we compare the density profiles of nitrogen and methane for $P=1.5$~atm and $P=3.0$~atm and temperature $T=90$~K. A comparison of the density profiles of nitrogen and ethane for the same thermodynamic conditions is shown in ~Figs.~\ref{fig:fig4} (C) and (D). A number of observations are noteworthy here. While the interfacial density of adsorbed nitrogen increases with increasing pressure, the difference between the maximum density of nitrogen at the interface and the density in the liquid phase becomes increasingly small with increasing pressure for nitrogen-methane system in the range of pressure studied here. On the other hand for nitrogen-ethane system, the difference between the maximum density at the interface and the density in the liquid-phase remains large due to smaller solubility. Furthermore, the interfacial width of the adsorbed nitrogen does not change appreciably with pressure for both the nitrogen-methane and nitrogen-ethane systems. The surface adsorption of gases on the liquid-gas interface has been studied in a number of other systems~\cite{king1972adsorption,massoudi1975effect,masterton1963surface,Minkara:2018aa}. 
\begin{figure}
\centering
\includegraphics[width=0.98\linewidth]{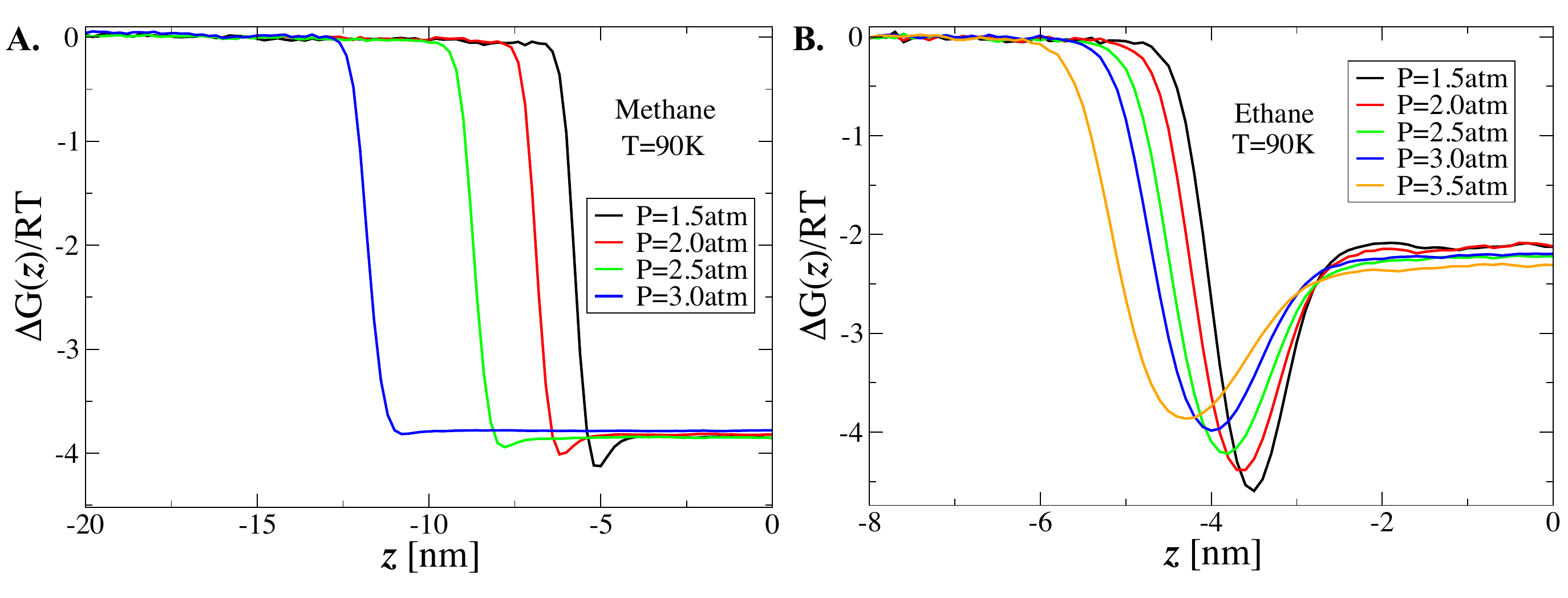}
\caption{Free energy profile, $\Delta G(z)$, at different temperatures for (A) nitrogen-methane and (B) nitrogen-ethane system. The data is only shown for the $z$-values close to the interface so that one can observe the vapor, the interface and the liquid regions.}
\label{fig:fig5}
\end{figure}
The partition coefficient, $K$, of two phases is defined as the ratio of the number density of the phases. Here we can define a $z$-dependent partition coefficient, $K(z)$ 
\begin{equation}
K(z) = \frac{\rho_N(z)}{\rho_N^v}
\end{equation}
where $\rho_N(z)$ is the density of nitrogen along the $z$-direction and $\rho_N^v$ is the density of the nitrogen in the vapor phase. Consequently, one can define excess free energy, $\Delta G(z)$, over the free energy of the vapor-phase as
\begin{equation}
\Delta G(z) = -RT\log K(z)
\end{equation}
where $R$ is the universal gas constant and $T$ is the temperature. In Figs.~\ref{fig:fig5}, we show $\Delta G(z)$ for different pressures at $T=90$~K for nitrogen-methane and nitrogen-ethane systems We find that $\Delta G(z)$ decreases upon entering the adsorbate region, reaches a minimum, and increases further and levels off in the liquid region, suggesting the free energy change in adsorbate region is larger compared to that in the liquid-region and hence adsorption at the interface. Moreover, we find that the magnitude of the free energy minimum, $|\Delta G_{\rm min}(z)|$, decreases with increasing pressure even though the effective density of nitrogen in the adsorbate region increases with increasing pressure. The decrease of $|\Delta G_{\rm min}(z)|$ with pressure arises because of relatively larger increase of the value of vapor-phase density of nitrogen with pressure. Furthermore, we find that the difference between the free energy of the adsorbate region and the liquid phase increases upon decreasing pressure. To find the free energy associated with the exchange with the adsorbate region, $\Delta G_{\rm ads}$, one must define the effective density of nitrogen, $\bar{\rho_N^I}$, in the adsorbate region. We define the adsorbate region as the region between the vapor and liquid phases in which the derivative of nitrogen-density is non-zero. This definition of characterizing interface, gas, and liquid regions are adopted from Ref.~\cite{buldyrev2007water}. $\Delta G_{\rm ads}$ is defined as
\begin{figure}
\centering
\includegraphics[width=0.7\linewidth]{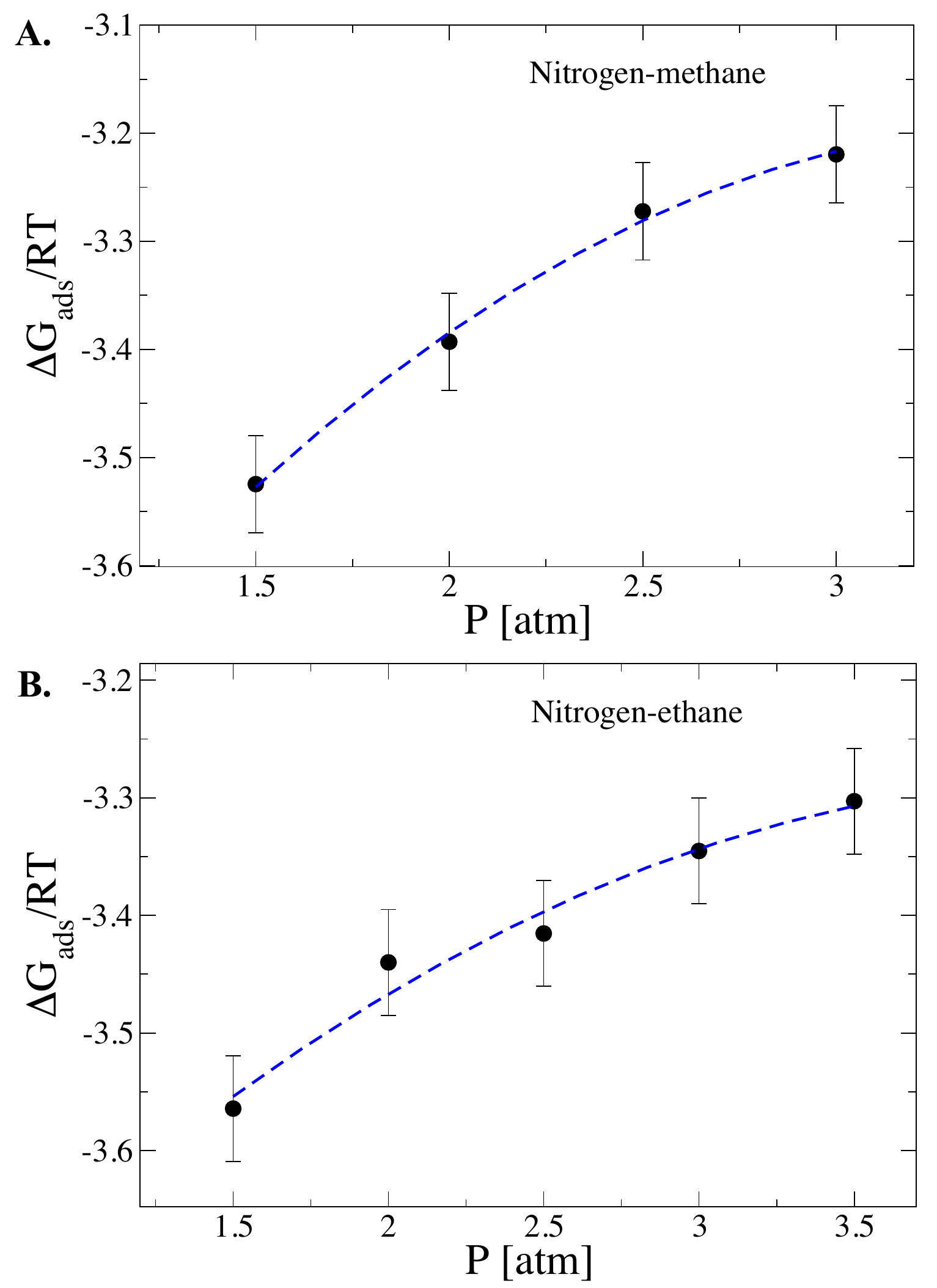}
\caption{Pressure dependence of the free energy of adsorption, $\Delta G_{\rm{ads}}$, for (A) nitrogen-methane, and (B) nitrogen-ethane systems, respectively. Dotted lines are the guide to the eye. The error bar on $\Delta G_{\rm{ads}}$ is estimated from four independent simulations of nitrogen-methane system at $T=90$~K and $P=1.5$~atm. The same error is assumed for all the state points studied here. $\Delta G_{\rm{ads}}$ decreases slightly with increasing pressure.}
\label{fig:fig6}
\end{figure}
\begin{equation}
\Delta G_{\rm ads} = -RT \log{\frac{\bar{\rho_N^I}}{\rho_N^{v}}}
\end{equation}
 In Figs.~\ref{fig:fig6}, we show $\Delta G_{\rm ads}$  as a function of pressure at $T=90$~K for nitrogen-methane and nitrogen-ethane systems, respectively. To estimate the error on the values of $\Delta G_{\rm ads}$, we performed four independent simulations of nitrogen-methane binary system at $T=90$~K and $P=1.5$~atm. The error is estimated as the standard deviation of the values obtained in these simulations. Same error was assumed for all other state points studied here. We find that $\Delta G_{\rm ads}$ values for both nitrogen-methane and nitrogen-ethane are similar and increase weakly with pressure.
\subsection*{Dissolution kinetics at the interface}
\begin{figure}
\centering
\includegraphics[width=0.98\linewidth]{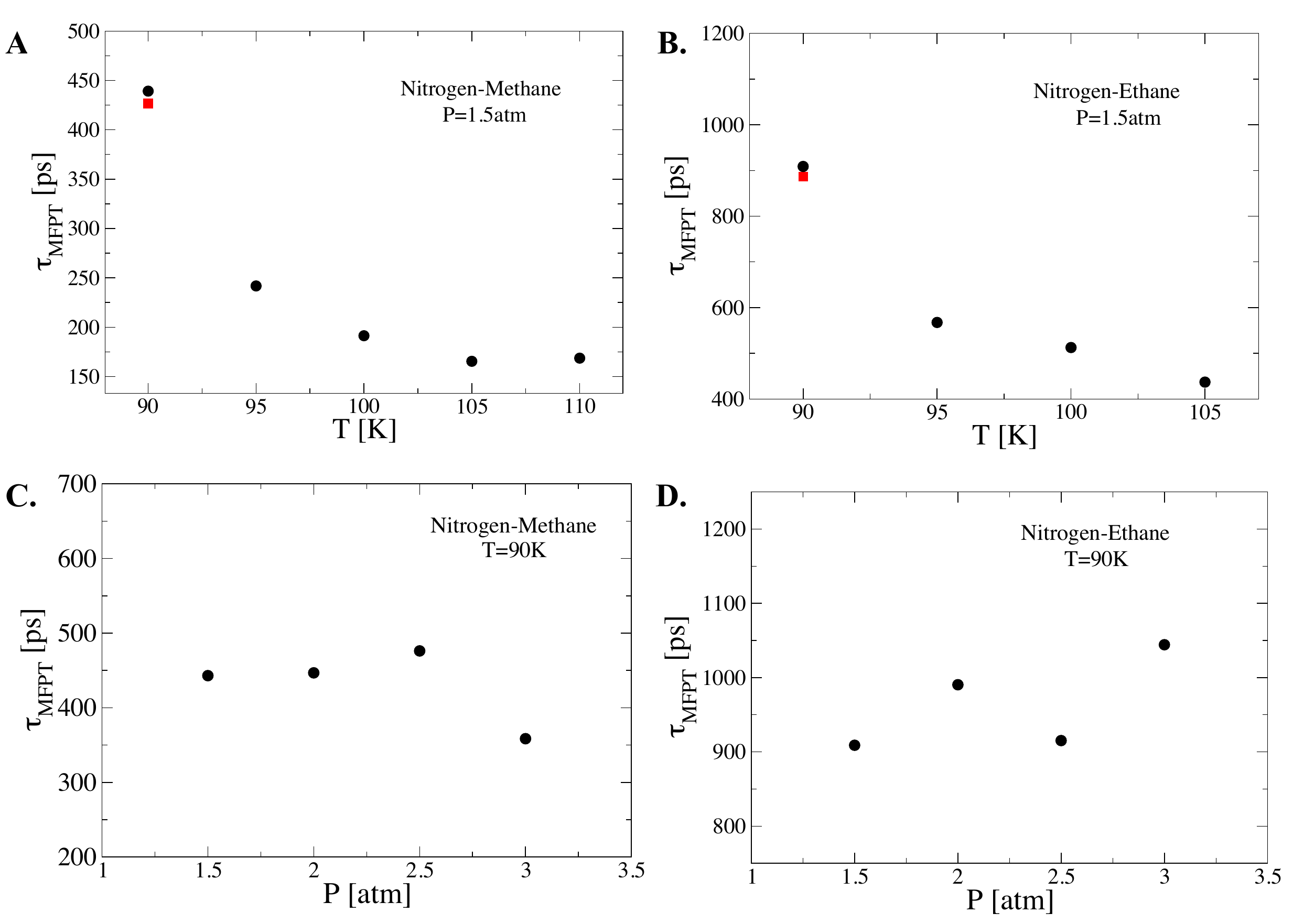}
\caption{Temperature dependence of mean first passage time, $\tau_{\rm{MFPT}}$ for (A) Nitrogen-methane, and (B) Nitrogen-ethane systems, respectively for $P=1.5$~atm. $\tau_{\rm{MFPT}}$ decreases upon increasing temperature. Black solid circles are the values calculated from molecular dynamics simulations and red solid squares are the values calculated using the diffusion model in Eq.~\ref{eq:eqBD} for $P=1.5$~atm and $T=90$~K. We find excellent agreement between $\tau_{\rm MFPT}$ calculated from molecular dynamics simulation and the values using Eq.~\ref{eq:eqBD}. Pressure dependence of $\tau_{\rm MFPT}$ at $T=90$~K for (C) nitrogen-methane, and (D) nitrogen-ethane systems, respectively. $\tau_{\rm MFPT}$ does not change appreciably with pressure.}
\label{fig:fig7}
\end{figure}

We next address the question of how a nitrogen molecule enters into the liquid phase from the surface. In order to investigate the absorption of a nitrogen molecule into the liquid phase from the interface, we have computed the mean first-passage time, $\tau_{\rm {MFPT}}$, that a nitrogen molecule starting at the position of the free energy minimum in Fig~\ref{fig:fig5} at the interface takes before it enters the liquid phase. In Figs.~\ref{fig:fig7}(A) and (B), we show the temperature dependence of $\tau_{\rm{MFPT}}$ for $P=1.5$~atm for nitrogen-methane and nitrogen-ethane systems, respectively. $\tau_{\rm{MFPT}}$ decreases with increasing temperature for both the systems. Furthermore, we find that $\tau_{\rm{MFPT}}$ is larger for nitrogen-ethane interface compared to nitrogen-methane interface. Figures~\ref{fig:fig7}(C) and (D) show the pressure dependence of $\tau_{\rm{MFPT}}$ for $T=90$~K for nitrogen-methane and nitrogen-ethane interface, respectively. $\tau_{\rm{MFPT}}$ does not show appreciable changes with pressure. An important question arises whether the absorption of a nitrogen molecule from interface is purely diffusive or there is an additional energy barrier of solvation of nitrogen that plays a role in the relaxation of hydrocarbon-nitrogen interface. To assess this, we use Langevin equation~\cite{gardiner2009stochastic,van1992stochastic} to calculate the mean first-passage time (MFPT) of escape of a nitrogen molecule from the minimum of the free energy to the liquid phase. Assuming an one-dimensional diffusion along $z$-direction on an underlying free energy surface, $\Delta G(z)$, the dynamics of  $z$ is governed by
\begin{align}
\dot{z} &= v_z \\
m\dot{v_z} &= -\xi v_z - \partial_z\Delta G(z) +f(t)
\end{align}
where $z$ is the position of a nitrogen molecule, $v_z$ is the velocity, $\xi$ is the friction coefficient,  and $f(t)$ is a delta-function correlated thermal noise with $\left<f(t)f(0)\right>=2\xi k_BT\delta(t)$ and zero mean, $\left< f(t)\right>=0$, where $T$ is the temperature, and $k_B$ is the Boltzmann constant. In the high friction limit, the above set of equations reduces to
\begin{equation}
\dot{z} = - \frac{D}{k_BT}\partial_z\Delta G(z) +f'(t)
\label{eq:eqBD}
\end{equation}
where $D$ is the diffusion coefficient of nitrogen, $\left< f'(t)\right >=0$ and $\left<f'(t)f'(0)\right>=2D\delta(t)$. In principle, $D$ is a function of position. For the sake of simplicity, we assume a constant diffusion across the interface in an effective density, $\bar{\rho}$, of the liquid. In separate simulations, we have computed the diffusion coefficient, D, of nitrogen for nitrogen+methane and nitrogen+ethane binary systems with effective $\bar{\rho}=\bar{\rho}_N+\bar{\rho}_{M/E}$, where $\bar{\rho}_N$ and $\bar{\rho}_{M/E}$ are the densities of nitrogen and methane/ethane in the liquid phase, respectively. We compute the value of $D$ for nitrogen-methane and nitrogen-ethane systems at  $T=90$~K and at their corresponding liquid-phase composition at $P=1.5$~atm. We find that the values of D is $3.03\rm{x}10^{-5}$~cm$^2$/s for nitrogen-methane mixture and $D$ is $0.822\rm{x}10^{-5}$~cm$^2$/s for nitrogen-ethane mixture. With the values of diffusion constant at our disposal, we next performed stochastic simulations of Eq.~\ref{eq:eqBD} and calculated the MFPT  for the escape of a nitrogen molecule from the minimum of free energy into the liquid phase. The MFPT was computed as the average of first-passage time over $10^3$ realizations. In Figure~\ref{fig:fig7}, we show these values as red solid squares.  We find that the values of $\tau_{\rm MFPT}$ estimated from Eq.~\ref{eq:eqBD} are in good agreement with molecular dynamics simulation. The slight discrepancy may arise because of the assumption of a uniform diffusion constant across the interface. These results suggest that once the nitrogen molecule is adsorbed on the surface they diffuse into the liquid phase without any additional energy barrier. Therefore, the time-scale of the diffusive dynamics across the interface is the bottleneck for interfacial relaxation. 
\section*{Summary and Discussion}

We have studied the pressure and temperature dependence of the solubility and surface adsorption of nitrogen in methane and ethane by performing extensive vapor-liquid equilibrium simulations of binary mixtures for range of pressure between $1.5$~atm and $3.5$~atm and temperature between $90$K and $110$K, thermodynamic conditions that may exist on the Saturn's giant moon, Titan. We find that the solubility of nitrogen in both methane and ethane increases with increasing pressure at a fixed temperature of $90$~K. The pressure behavior of the solubility of nitrogen in methane exhibit a linear dependence while a exponential pressure dependence of the solubility of nitrogen in ethane is found at $T=90$~K. Solubility of nitrogen in both methane and ethane exhibits exponential decrease with temperature at a fixed pressure of $P=3.0$~atm. The solubility of nitrogen in methane is much larger compared to that in ethane in the range of pressure and temperature studied here. Our results are in quantitative agreement with the available experimental measurements of solubility of nitrogen in methane and ethane. Furthermore, we find that the excess surface adsorption of nitrogen over the liquid concentration decreases with increasing pressure for $T=90$~K while the adsorption free energy increases slightly with increasing pressure. Moreover, we find that the surface tension decreases linearly with pressure for both nitrogen-methane and nitrogen-ethane systems at $T=90$~K. The rate of decrease of surface tension with pressure for nitrogen-ethane system is much larger, almost about twice as large, compared to the nitrogen-methane system. The values of surface tension rule out the likelihood of homogeneous nucleation of nitrogen bubbles in the Titan's liquid hydrocarbon bodies. Heterogeneous nucleation on the sea bed or near suspended particles in the seas could be other plausible scenario of bubble formation in Titan's seas as suggested by Cordier et. al.~\cite{cordier2018bubbles} and should be investigated in future. We have also investigated the dissolution kinetics of at the liquid-gas interface. We looked at if the absorption kinetics of nitrogen molecules from the interface into the liquid region is purely diffusive or there are additional energy barriers. We find that the absorption of a nitrogen molecule from the interface into the liquid-phase is diffusive and does not involve any appreciable energy barrier. Specifically, we find that the mean first-passage time of escape of a nitrogen molecule from the position of minimum in the free energy into the liquid-phase calculated from the molecular dynamics simulations agrees well with a pure diffusion model of nitrogen on the underlying free energy curve. The mean first-passage time is less than a $1$~ns for all the pressure and temperature investigate here. This suggests that the time-scale of liquid-gas interfacial relaxation is solely dictated by the diffusion of nitrogen molecules. 
\section*{Methods}
\begin{figure}
\centering
\includegraphics[width=.8\linewidth]{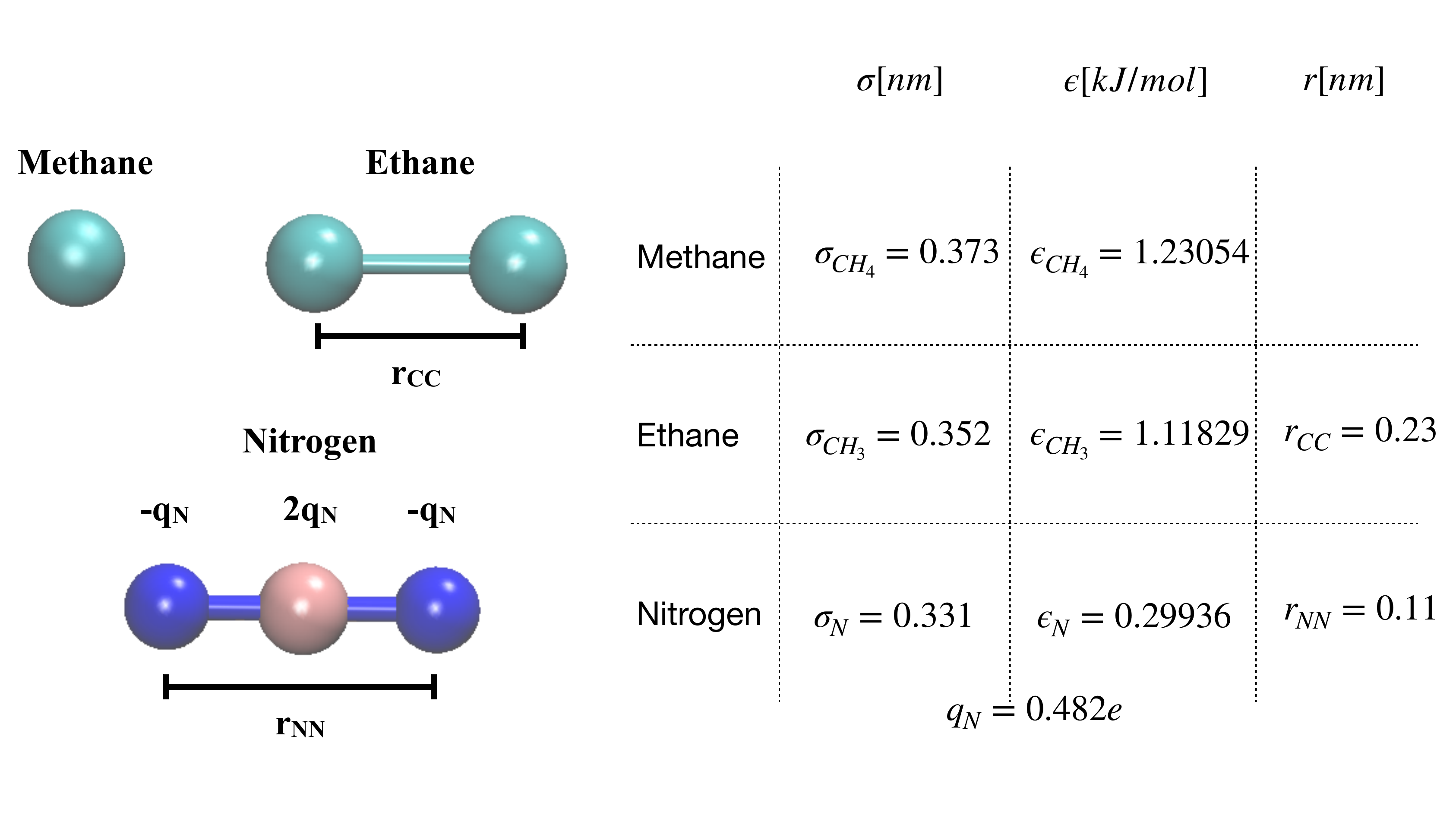}
\caption{TRAPPE-UA(2) models of methane, ethane, and nitrogen. The parameters of the models are also listed. Note that an improved parametrization of ethane, trappe-UA2, is used.}
\label{fig:fig8}
\end{figure}
Vapor-liquid equilibrium simulations (VLE) is a well established method to study the phase-equilibria of liquid and vapor phases. Specifically, we have performed extensive VLE simulations of (i) binary mixtures of nitrogen and methane, (ii) nitrogen and ethane for a range of pressure between $1.5$~atm and $3.5$~atm and temperatures between $90$~K and $110$~K.  Simulations were performed in Gromacs4.6.5~\cite{BERENDSEN199543,Lindahl:2001aa,Van-Der-Spoel:2005aa}. The trappe-UA force field~\cite{keasler2012transferable,wick2005transferable,wick2000transferable,martin1999novel} was used to model methane, and ethane was modeled using an improved parameterization of ethane, trappe-UA2~\cite{shah2017transferable}. In the trappe-UA force field, the methane is represented as an atom and for ethane, each $CH_3$ groups is represented by one atom (see Figure~\ref{fig:fig8}). The distance between the coarse-grained $CH_3$ atom is $r_{\rm{CC}}=0.23$~nm. Trappe-small parameterization was used to model nitrogen as a three-site model with nitrogens carrying negative partial charges, $q_N=-0.482e$, and a virtual site that sits at the center of mass of the molecule carries a positive partial charge of $2q_{N}=0.964e$~\cite{keasler2012transferable} (see Figure~\ref{fig:fig1}). The distance between the two nitrogen atoms is $r_{\rm{NN}}=0.11$~nm. The van der Waals radii for $CH_4$, $CH_3$, and $N$ are $\sigma_{CH_4}=0.373$~nm, $\sigma_{CH_3}=0.352$~nm, and $\sigma_{N}=0.351$~nm, respectively. The van der Waals interaction energy  for $CH_4$, $CH_3$, and $N$ are $\epsilon_{CH_4}=0.123054$~kJ/mol, $\epsilon_{CH_3}=0.0.11829$~kJ/mol, and $\epsilon_{N}=0.29936$~kJ/mol, respectively. The short-range van der Waals interaction potential between two atoms $i$ and $j$ is given by
\begin{equation}
U(r_{ij}) = 4\pi \epsilon_{ij} \left [\left(\frac{r_{ij}}{\sigma_{ij}}\right)^{12}-\left(\frac{r_{ij}}{\sigma_{ij}}\right)^6\right]
\end{equation}
Lorentz-Berthelot~\cite{frenkel2001understanding} rule was used to model the cross-interactions with $\sigma_{ij}=\frac{\sigma_{i}+\sigma_j}{2}$ and $\epsilon_{ij}=\sqrt{\epsilon_{i}\epsilon_j}$. The short-range interactions were treated with a  cut-off of $1.5$~nm and particle-mesh-Ewald(PME)~\cite{frenkel2001understanding} was used for the long range interactions.  Since the system is not homogeneous, dispersion corrections to the energy and pressure were not applied. The initial configurations for different pressures and temperatures of the binary systems were created in two steps. First, a liquid-phase simulations of hydrocarbons (methane/ethane) were performed at the corresponding pressure and temperature with fixed simulation box sizes $L_x=5$~nm and $L_y=5$~nm along $x$ and $y$-directions, respectively. Separately, a simulation of nitrogen at a fixed volume with $L_x=L_y=5$~nm and with appropriate number of nitrogen molecules was performed at the same pressure and temperature. Finally, two simulation boxes of nitrogen and hydrocarbon was put together to form a simulation box consisting of nitrogen and hydrocarbon. A typical configuration of the nitrogen-ethane binary system is shown in Figure 1. The dimension of the final simulation box was $L_x=L_y=5.0$nm$<< L_z$. In such a box, the interface is stable and forms along the smallest surface area in the $xy$-plane, perpendicular to the long-axis. The number of molecules of methane and ethane was fixed to $3000$ and $2000$ for all the binary mixture simulations and the number for nitrogen molecules varied for different pressures and temperatures depending on the solubility and the gas phase density. The equations of motion are integrated with a time step of $2$~fs and velocity rescaling is  used to attain constant temperature and anisotropic Berendsen barostat for constant pressure in the $z$-direction.  After the equilibration for $80$~ns, we ran the simulations for additional $80$~ns for each state point  and the equilibrium averages are calculated from these trajectories. To check if the system is well equilibrated, we monitored the potential energy of the system throughout the production run, and we did not find any drift for all the simulations performed here. 

\medskip

\noindent{\bf \large Author Contributions:} P.K. designed and performed the research, analyzed the data and wrote the paper. P.K. and V.F.C read and revised the paper.
\medskip
\begin{acknowledgments}
Authors would like to thank University of Arkansas High Performance Computing Center for providing computational time. V. F. Chevrier acknowledges funding from NASA Cassini Data Analysis Program grant no. NNX15AL48G.
\end{acknowledgments}

\bibliographystyle{apsrev4-1}
\bibliography{kumar2020titan}

\end{document}